\documentstyle[preprint,aps]{revtex}
\begin{document}
\setlength{\baselineskip}{18pt}
\begin{flushright}
CU-TP/98-05\\
May, 1998
\end{flushright}
\begin{center}
{\Large Atmospheric neutrino oscillations in three-flavor neutrinos}\\
\vspace{1cm}
T. Teshima\footnote{E-mail address: teshima@isc.chubu.ac.jp}
 and T. Sakai\\
\vspace{1cm}
Department of Applied Physics,  Chubu University, Kasugai 487-8501, Japan\\
\end{center}
\begin{abstract}
We analyzed the atmospheric neutrino experiments of SuperKamiokande including zenith angle dependence's using the three-flavor neutrino framework with the hierarchy $m^2_1\approx m^2_2\ll m^2_3$. Taking into account the  terrestrial, solar neutrino experimental data and the atmospheric neutrino experiments including the sub-GeV and multi-GeV data in SuperKamiokande, large angle solution in the solar neutrino experiments is favored and the range of the mass parameter $\Delta m^2_{23}$ is restricted between $0.08{\rm eV}^2 - 2{\rm eV}^2$. Allowed regions of mixing parameters are  $(\theta_{13}<4^\circ,\ 27^\circ<\theta_{23}<32^\circ)$ for $\Delta m_{23}^2=1{\rm eV}^2$ and $(\theta_{13}<3^\circ,\ 28^\circ<\theta_{23}<33^\circ)$ for $\Delta m_{23}^2=0.1{\rm eV}^2$.
\end{abstract}
\vspace{0.5cm}

\vspace{1cm}
\setlength{\baselineskip}{16pt}
\section{Introduction}
Observing the neutrino oscillations is one of the most important experiments with respect to new physics beyond the Standard Model. In recent SuperKamiokande atmospheric experiments\cite{SUPERKAMIOKANDE,KANEYUKI}, definite atmospheric neutrino anomaly is observed following that in Kamiokande experiments\cite{HIRATA,FUKUDA} and zenith angle dependence's of double ratios $R(\mu/e)$ in atmospheric neutrinos are obtained, those which have decreasing characters of double ratios to increasing of zenith angle for both sub-GeV and multi-GeV experiments. Decreasing character suggests the strong evidence for neutrino oscillations and observing these decreasing characters in both sub-GeV and multi-GeV experiments give definite information about the neutrino mixing parameters; $\Delta m^2$ and $\sin^22\theta$.
\par 
In the terrestrial neutrino experiments, many experimental groups have tried to observe the neutrino oscillations. The E531\cite{E531}, CHORUS and NOMAD\cite{CHORUS} experiments using the beam from accelerators search for $\nu_\tau$ appearance in $\nu_\mu$, and E776\cite{E776}, KARMEN\cite{EMU} and LSND\cite{LSND} experiments using the accelerator beams are searching for $\nu_\mu \to \nu_e$ and $\bar{\nu}_\mu\to \bar{\nu}_e$ oscillations. The experiments using nuclear power reactor\cite{BUGEY,CHOOZ} search for the disappearance of $\bar{\nu}_e$, in which $\bar{\nu}_e\to\bar{\nu}_X (X=\mu, \tau)$ transitions are expected. These experiments do not observe significantly large neutrino transitions. Especially, the recent Chooz experiment\cite{CHOOZ} excludes the parameter region given approximately by $\Delta m^2 > 0.9\times10^{-3}{\rm eV}^2$ for maximum mixing and $\sin^22\theta>0.18$ for large $\Delta m^2$.         
\par 
In this paper, we analyze the atmospheric neutrino experimental data obtained in SuperKamiokande including the zenith angle dependence in three-flavor neutrino framework with a hierarchy of neutrino masses $m_1\approx m_2\ll m_3$. In previous work\cite{TESHIMA}, we have analyzed the solar neutrino, terrestrial neutrino and atmospheric neutrino experiments in three-flavor neutrino framework with hierarchy  $m_1\approx m_2\ll m_3$ including the zenith angle dependence's partially. In this work, we perform the full analysis of zenith angle dependence's in SuperKamiokande atmospheric neutrino experiment\cite{KANEYUKI} and analysis of Chooz data\cite{CHOOZ} in terrestrial experiment.
\par
In three-flavor neutrino framework with hierarchy $m_1\approx m_2\ll m_3$, there are 5 parameters $\sin^22\theta_{12},\ \theta_{13},\ \theta_{23},\ \Delta m^2_{12}=m_2^2-m_1^2,\ \Delta m^2_{23}=m_3^2-m_2^2 $ concerned with the neutrino oscillation. For solar neutrino deficit, MSW solution\cite{KUO,FOGLIS,TESHIMA} considering the matter effects predicts the large angle solution $\sin^22\theta_{12}=0.6-0.9$, $\Delta m^2_{12}=4\times10^{-6}-7\times10^{-5}{\rm eV}^2$ and small angle solution $\sin^22\theta_{12}=0.003-0.01$, $\Delta m^2_{12}=3\times10^{-6}-1.2\times10^{-5}{\rm eV}^2$ for $\theta_{13}=0^\circ-20^\circ$, and these large angle and small angle solution are merged for  $\theta_{13}=25^\circ-50^\circ$. The vacuum solution in solar neutrino is obtained as $\Delta m^2_{12}\sim10^{-10}{\rm eV}^2$\cite{VACUUME}. In terrestrial neutrino experiments, Fogli {\it et al.}\cite{FOGLIT} have analyzed thoroughly and obtained the allowed regions on $\tan\theta^2_{13}-\tan^2\theta_{23}$ plane for various values of mass parameter $\Delta m^2_{23}$. We also analyzed the terrestrial neutrino experiments\cite{TESHIMA} in which we have not included Chooz experiment, and obtained the similar results as the one of Fogli {\it et al.} Obtained allowed region of $\tan^2\theta_{13}-\tan^2\theta_{23}$ plane, for example for $\Delta m^2_{23}=0.02{\rm eV}^2$, is $\tan^2\theta_{13}>25,  \tan^2\theta_{13}<0.04$ for all values of $\tan^2\theta_{23}$. For higher values of $\Delta m^2_{23}$, the allowed region on $\tan^2\theta_{13}-\tan^2\theta_{23}$ plane was more restricted, and for lower values of it, the allowed region spread out and for $\Delta m^2_{23}<0.01{\rm eV}^2$ there was no restriction on $\tan^2\theta_{13}-\tan^2\theta_{23}$ plane.
\par 
Atmospheric neutrino experiments have been analyzed in three-flavor neutrino framework with neutrino mass hierarchy $m_1\approx m_2\ll m_3$ by many authors\cite{FOGLIA,THREE,TESHIMA}. The atmospheric neutrino anomaly of Kamiokande experiments\cite{KAMIOKANDE2} has been explained by $\nu_\mu \leftrightarrow \nu_\tau$ oscillation with a parameter $\Delta m^2_{23} \sim 10^{-2}-10^{-1}{\rm eV}^2$. These discussions contain the zenith angle dependence's of sub-GeV and/or multi-GeV Kamiokande data. The zenith angle dependence's in recent SuperKamiokande atmospheric experiments\cite{SUPERKAMIOKANDE,KANEYUKI} have decreasing character not only for multi-GeV experiment but also for sub-GeV one. We have analyzed these dependence's partially\cite{TESHIMA} and obtained a result that large angle solution for $\sin^22\theta_{12}$ is favorable. In this paper, we will analyze these dependence's thoroughly.

\section{Neutrino oscillation in three-flavor neutrinos}
Weak currents for the interactions producing and absorbing neutrinos are described as 
\begin{equation}
J_{\mu}=2\sum^3_{\alpha,\beta=1}\bar{l}_{L\alpha}\gamma_{\mu}U_{l_{\alpha}\beta}\nu_{L\beta},
\end{equation}
where $l_{\alpha}\ (l_1=e,\ l_2=\mu,\ l_3=\tau)$ represents the lepton flavor, $\nu_{\beta}$ the neutrino mass eigenstate and $U$ is the lepton mixing matrix.
$U$ is the unitary matrix corresponding to the CKM matrix $V_{CKM}^{\dagger}$ for quarks defined by
\begin{equation}
U=U_lU^{\dagger}_{\nu},
\end{equation}
where the unitary matrices $U_l$ and $U_{\nu}$ transform mass matrices $M^l$ for charged leptons and $M^{\nu}$ for neutrinos to diagonal mass matrices as
\begin{equation}
\begin{array}{l}
U_lM^lU^{-1}_l={\rm diag}[m_e, m_{\mu}, m_{\tau}], \\
U_{\nu}M^{\nu}U^{-1}_{\nu}={\rm diag}[m_1, m_2, m_3].
\end{array}
\end{equation}
We present the unitary matrix neglecting the {\it CP} violation phases as 
\begin{eqnarray}
 U&=&e^{i\theta_{23}\lambda_7}e^{i\theta_{13}\lambda_5}e^{i\theta_{12}
             \lambda_2}\nonumber\\
            &=&\left(
      \begin{array}{ccc}
      c_{12}^{\nu}c_{13}^{\nu} & s_{12}^{\nu}c_{13}^{\nu} & s_{13}^{\nu} \\
      -s_{12}^{\nu}c_{23}^{\nu}-c_{12}^{\nu}s_{23}^{\nu}s_{13}^{\nu} & c_{12}^{\nu}c_{23}^{\nu}-s_{12}^{\nu}s_{23}^{\nu}s_{13}^{\nu} & s_{23}^{\nu}c_{13}^{\nu} \\
      s_{12}^{\nu}s_{23}^{\nu}-c_{12}^{\nu}c_{23}^{\nu}s_{13}^{\nu} & -c_{12}^{\nu}s_{23}^{\nu}-s_{12}^{\nu}c_{23}^{\nu}s_{13}^{\nu} & c_{23}^{\nu}c_{13}^{\nu} 
      \end{array}\right), \\
 & & \quad\quad c_{ij}^{\nu}=\cos{\theta}^{\nu}_{ij},\ \  s_{ij}^{\nu}=\sin{\theta}^{\nu}_{ij}, \nonumber 
\end{eqnarray}  
where $\lambda_i$'s are Gell-Mann matrices.
\par 
The probabilities for transitions  $\nu_{l_\alpha} \to \nu_{l_\beta}$ are written as 
\begin{eqnarray}
P(\nu_{l_\alpha}\to\nu_{l_\beta})&&=|<\nu_{l_\beta}(t)|\nu_{l_\alpha}(0)>|^2 = \delta_{l_\alpha l_\beta}+p_{\nu_{l_\alpha}\to\nu_{l_\beta}}^{12}S_{12}+p_{\nu_{l_\alpha}\to\nu_{l_\beta}}^{23}S_{23}+p_{\nu_{l_\alpha}\to\nu_{l_\beta}}^{31}S_{31},\nonumber \\
&&p_{\nu_{l_\alpha}\to\nu_{l_\beta}}^{12}=-2\delta_{l_\alpha l_\beta}(1-2U_{l_\alpha3}^2)+2(U_{l_\alpha1}^2U_{l_\beta1}^2+U_{l_\alpha2}^2U_{l_\beta2}^2-U_{l_\alpha3}^2U_{l_\beta3}^2), \nonumber \\       
&&p_{\nu_{l_\alpha}\to\nu_{l_\beta}}^{23}=-2\delta_{l_\alpha l_\beta}(1-2U_{l_\alpha1}^2)+2(-U_{l_\alpha1}^2U_{l_\beta1}^2+U_{l_\alpha2}^2U_{l_\beta2}^2+U_{l_\alpha3}^2U_{l_\beta3}^2), \nonumber \\
&&p_{\nu_{l_\alpha}\to\nu_{l_\beta}}^{31}=-2\delta_{l_\alpha l_\beta}(1-2U_{l_\alpha2}^2)+2(U_{l_\alpha1}^2U_{l_\beta1}^2-U_{l_\alpha2}^2U_{l_\beta2}^2+U_{l_\alpha3}^2U_{l_\beta3}^2) ,\nonumber \\
&&
\end{eqnarray}
where $S_{ij}$ is the term representing the neutrino oscillation;  
\begin{equation}
S_{ij}=\sin^21.27\frac{\Delta m^2_{ij}}{E}L,
\end{equation}
in which $\Delta m^2_{ij}=|m^2_i-m^2_j|$, $E$ and $L$  are measured in units eV$^2$, GeV and km, respectively.
\par
The values of neutrino masses are not known precisely, but two mass parameters are necessary to account for the solar neutrino experiments, in which  mass parameter $\Delta m^2$ is $10^{-4} - 10^{-5}{\rm eV}^2$ or $\sim 10^{-10}{\rm eV}^2$ is obtained \cite{KAMIOKANDE2}, and  for the atmospheric experiments, in which $\Delta m^2$ is  $10^{-1} - 10^{-2}{\rm eV}^2$ \cite{KAMIOKANDE2}. Then it seems that the most economical mass hierarchy is that lower two neutrino masses in three neutrinos are very close and another one is rather far away from them. Then we assume that three neutrino masses have such a mass hierarchy as  
\begin{equation}
m_1 \approx m_2 \ll m_3. \label{masshie}
\end{equation}
In this mass hierarchy Eq.~(\ref{masshie}), $\Delta m_{12}^2\ll\Delta m_{23}^2 \simeq \Delta m_{13}^2$, the expression Eq. (5) for the transition probabilities $P(\nu_{l_{\alpha}}\to\nu_{l_{\beta}})$ are rewritten as
\begin{eqnarray}
P(\nu_e\to\nu_e)&=&1-2(1-2U_{e3}^2-U_{e1}^4-U_{e2}^4+U_{e3}^4)S_{12}-4U_{e3}^2(1-U_{e3}^2)S_{23}, \nonumber \\
P(\nu_{\mu}\to\nu_{\mu})&=&1-2(1-2U_{\mu3}^2-U_{\mu1}^4-U_{\mu2}^4+U_{\mu3}^4)S_{12}-4U_{\mu3}^2(1-U_{\mu3}^2)S_{23}, \nonumber \\
P(\nu_\tau\to\nu_\tau)&=&1-2(1-2U_{\tau3}^2-U_{\tau1}^4-U_{\tau2}^4+U_{\tau3}^4)S_{12}-4U_{\tau3}^2(1-U_{\tau3}^2)S_{23}, \nonumber \\   
P(\nu_\mu\to\nu_e)&=&P(\nu_e\to\nu_\mu)=2(U_{\mu1}^2U_{e1}^2+U_{\mu2}^2U_{e2}^2-U_{\mu3}^2U_{e3}^2)S_{12}+4U_{e3}^2U_{\mu3}^2S_{23}, \nonumber \\
P(\nu_\tau\to\nu_e)&=&P(\nu_e\to\nu_\tau)=2(U_{\tau1}^2U_{e1}^2+U_{\tau2}^2U_{e2}^2-U_{\tau3}^2U_{e3}^2)S_{12}+4U_{e3}^2U_{\tau3}^2S_{23}, \nonumber \\
P(\nu_\tau\to\nu_\mu)&=&P(\nu_\mu\to\nu_\tau)=2(U_{\tau1}^2U_{\mu1}^2+U_{\tau2}^2U_{\mu2}^2-U_{\tau3}^2U_{\mu3}^2)S_{12}+4U_{\mu3}^2U_{\tau3}^2S_{23}.\nonumber \\
&& 
\label{trans.prob.}
\end{eqnarray}

\section{Numerical analyses of neutrino oscillations}
\subsection{Atmospheric neutrinos}
\par
The evidence for an anomaly in atmospheric neutrino experiments was pointed out  by Kamiokande Collaboration \cite{HIRATA,FUKUDA} and IMB Collaboration \cite{IMB} using the water-Cherencov experiments. More recently, SuperKamiokande Collaboration \cite{SUPERKAMIOKANDE,KANEYUKI} reports the more precise results on anomaly in atmospheric neutrino. The double ratios  $R(\mu/e)\equiv{R_{\rm expt}(\mu/e)}/{R_{\rm MC}(\mu/e)}$ obtained are summarized as 
\begin{mathletters}
\begin{eqnarray}
R(\mu/e)&=& \left\{
\begin{array}{l}
 0.60\mbox{\small${+0.07\atop-0.06}$}\pm0.05 \ \  \makebox{for Kamiokande (sub-GeV)\cite{HIRATA}},  \\
 0.57\mbox{\small${+0.08\atop-0.07}$}\pm0.07 \ \  \makebox{for Kamiokande (multi-GeV)\cite{FUKUDA}}, 
\end{array} \right. \\ 
 R(\mu/e)&=& 0.54\pm0.02\pm0.07 \ \ \makebox{for IMB \cite{IMB}},  \\
R(\mu/e)&=& \left\{
\begin{array}{l}
0.63\pm0.03\pm0.05 \ \ \makebox{for SuperKamiokande (sub-GeV)\cite{KANEYUKI}},  \\
0.60\pm0.06\pm0.07 \ \ {\rm for\ SuperKamiokande\ (multi\makebox{-}GeV) \cite{KANEYUKI}}.
\end{array}\right.      
\end{eqnarray}
\end{mathletters}
In the above data, sub-GeV experiments detect the visible-energy less than 1.33GeV. The zenith angle dependence's of double ratio $R(\mu/e)$ obtained in SuperKamiokande are shown in Fig. 1\cite{KANEYUKI}. The sub-GeV data has decreasing character to increasing of zenith angle in contrast to the zenith angle dependence of Kamiokande data which has no decreasing character\cite{HIRATA}. The multi-GeV data has also decreasing character to increasing of zenith angle similar to that of Kamiokande\cite{FUKUDA}. 
\par
The ratios ${R_{\rm expt}(\mu/e)}$ and ${R_{\rm MC}(\mu/e)}$ are defined as
\begin{mathletters}
\begin{eqnarray}
R_{\rm expt}(\mu/e)&=&\frac{\sum_{\alpha}\int\epsilon_{\mu}(E_{\mu})\sigma_{\mu}(E_{\nu},E_{\mu})F_{\alpha}(E_{\nu},\theta)P(\nu_\alpha\to\nu_\mu)dE_{\mu}dE_{\nu}d\theta}
        {\sum_{\alpha}\int\epsilon_{e}(E_{e})\sigma_{e}(E_{\nu},E_{e})F_{\alpha}(E_{\nu},\theta)P(\nu_\alpha\to\nu_e)dE_{e}dE_{\nu}d\theta}, \\
R_{\rm MC}(\mu/e)&=&\frac{\int\epsilon_{\mu}(E_{\mu})\sigma_{\mu}(E_{\nu},E_{\mu})F_{\mu}(E_{\nu},\theta)dE_{\mu}dE_{\nu}d\theta}
          {\int\epsilon_{e}(E_{e})\sigma_{e}(E_{\nu},E_{e})F_{e}(E_{\nu},\theta)dE_{e}dE_{\nu}d\theta},
\end{eqnarray}
\end{mathletters}
where the summation $\sum_\alpha$ are taken in $\mu$, e neutrino, and in these expressions antineutrino processes are contained. $\epsilon_{\beta}(E_{\beta})$ is the detection efficiency of the detector for $\beta$-type charged lepton with energy $E_{\beta}$, $\sigma_{\beta}$ is the differential cross section of $\nu_{\beta}$ and $F_{\alpha}(E_{\nu},\theta)$ is the incident $\nu_{\alpha}$ flux with energy $E_{\nu}$ and zenith angle $\theta$. $P(\nu_\alpha\to\nu_\beta)$ is the transition probability Eq.~(8) and it depends on the energy $E_\nu$ and the distance $L$ which depends on zenith angle $\theta$ as $L=\sqrt{(r+h)^2-r^2\sin^2\theta}-r\cos\theta$, where $r$ is the radius of the Earth and $h$ is the altitude of production point of atmospheric neutrino. The zenith angle dependence's of double ratio are defined as the equation similar to the Eq. (10) except for the no integration on zenith angle $\theta$. 
\par
Although information of $F_\alpha(E_\nu,\theta)$ etc. are given in Refs. \cite{GAISSER,HONDA,AGRAWAL}, we use the MC predictions for $f_\alpha(E_\nu, \theta)\equiv\int\epsilon_{\mu(e)}(E_{\mu(e)})\sigma_{\mu(e)}(E_{\nu},E_{\mu(e)})F_{\alpha}(E_{\nu},\theta)dE_{\mu(e)}$ in Ref. \cite{HIRATA} for sub-GeV experiment and Ref. \cite{FUKUDA} for multi-GeV experiment. Explicit $E_\alpha$ dependence of $f_\alpha(E_\alpha,\theta_\alpha)$ are shown in Appendix B of Ref.\cite{TESHIMA}. Since $P(\nu_\alpha\to\nu_\mu)$ and $P(\nu_\alpha\to\nu_e)$ are the functions of $(\Delta m_{12}^2, \Delta m_{23}^2, \theta_{12}, \theta_{13}, \theta_{23}, L, E)$, the double ratio $R(\mu/e)$ which is integrated in neutrino energy $E$ and zenith angle $\theta$ (related to $L$) is a function of $(\Delta m_{12}^2, \Delta m_{23}^2, \theta_{12}, \theta_{13}, \theta_{23})$. The zenith angle dependence of double ratios is a function of $(\Delta m_{12}^2, \Delta m_{23}^2, \theta_{12}, \theta_{13}, \theta_{23}, \theta)$. 
\par
In previous paper\cite{TESHIMA}, we estimated the double ratio $R(\mu/e)$ fixing the parameters $(\Delta m_{12}^2, \sin^22\theta)$ which have been determined in solar neutrino experimental data\cite{SAGE,GALLEX,HOMESTAKE,KAMIOKANDES} using the three-flavor neutrino analyses\cite{FOGLIS,TESHIMA}, as  $\Delta m_{12}^2=3\times10^{-5}{\rm eV}^2$, $\sin^22\theta_{12}=0.7$ which corresponds to large angle solution and $\Delta m_{12}^2=10^{-5}{\rm eV}^2$, $\sin^22\theta_{12}=0.005$ which corresponds to small angle solution. We showed these plots in Fig. 2. Contour lines correspond to the upper and lower values of $R(\mu/e)$ in SuperKamiokande data. We showed the plots of sub-GeV experiment in Figs. 2(a)-(d) and plots of multi-GeV one in Figs. 2(e)-(f), and in these figures solid lines denote the large angle solution plots and dotted lines the small angle solution plots. In Figs. 2(e)-(h), dotted lines are close to solid lines. Same analysis has been performed by the third paper in Ref. \cite{FOGLIA} and our results obtained are similar to those.
\par
In Fig. \ref{fig3}, we showed the contour plots of $\chi^2$ taken from the zenith angle dependence data (values drawn from Fig. 1) for double ratio $R(\mu/e)$ on $\tan^2\theta_{13}-\tan^2\theta_{23}$ plane for various $\Delta m^2_{23}$. Contour curves denoted by dotted, thick solid and thin solid line correspond to the values of $\chi^2=7.78$ ($90\%$ C.L. for 4DF), 9.49 ($95\%$ C.L. for 4DF) and 13.3 ($99\%$ C.L. for 4DF), respectively. We showed the plots of sub-GeV experiment in Fig. \ref{fig3}(a)-(d) and plots of multi-GeV one in Fig. \ref{fig3}(e)-(h). These plots are taken for large angle solutions of $\sin^22\theta_{12}$ because  $\chi^2$ for small angle solution cannot be less than 12 in sub-GeV experiments. This is seen by the decreasing character of zenith angle dependence in sub-GeV data. We showed the predicted zenith angle dependence curves of $R(\mu/e)$ on the graph of zenith angle dependence data in Fig. 1. Fig. 1(a) represents the sub-GeV experimental case on parameters $(\Delta m^2_{23}=0.1{\rm eV}^2,\ \theta_{13}=2.4^\circ,\ \theta_{23}=25^\circ)$. In this figure, solid curve denotes the large angle solution case and dotted curve the small angle solution case. From this figure, we can understand the small angle solution case do not give the small $\chi^2$. It can be said that the decreasing character of sub-GeV data is caused by the large $\mu-e$ mixing. In Fig. 1(b), we showed the multi-GeV experimental case on parameters $(\Delta m^2_{23}=0.1{\rm eV}^2,\ \theta_{13}=4^\circ,\ \theta_{23}=45^\circ)$. In this case, there is no difference between large angle solution and small one. Decreasing character in multi-GeV data is explained by the large $\mu-\tau$ mixing.
\par
In Fig. 4, we showed the contour plots of $\chi^2$ of sub-GeV plus multi-GeV experiments. These are taken for large angle solution. Contour curves denoted by dotted , thick solid and thin solid line correspond to the values of $\chi^2$ 14.7(90$\%$ C.L. for 9DF),  16.9(95$\%$ C.L. for 9DF) and 21.7(99$\%$ C.L. for 9DF), respectively. Thus, if we claim that both sub-GeV data and multi-GeV one should be explained simultaneously, the allowed region on parameters become narrow. The best fit by $\chi^2$ for sub-GeV plus multi-GeV data is obtained on parameters $(\Delta m^2_{23}=0.1{\rm eV}^2,\ \theta_{13}=33^\circ,\ \theta_{23}=45^\circ)$ on which $\chi^2$ is $13.1$. Although the allowed region on parameters is obtained from the atmospheric neutrino experiments, all regions are not accepted because the terrestrial neutrino experiments seem unfavorable for the neutrino oscillation. In next subsection, we will discuss the terrestrial neutrino experiments, thoroughly.  

\subsection{Terrestrial neutrinos}
Although we have study the terrestrial neutrino experiments in previous paper\cite{TESHIMA}, we did not include the Chooz experiment\cite{CHOOZ}. We include it in this present analysis. In the short baseline experiments, there are E531 \cite{E531}, CHORUS and NOMAD \cite{CHORUS} accelerator experiments searching for $\nu_{\tau}$ appearance in $\nu_{\mu}$. We used the data of E531, CHORUS and NOMAD experiments;
\begin{eqnarray}
&&P(\nu_{\mu}\to\nu_{\tau})<2\times10^{-3}\ \ (90\%\ {\rm C.L.}),\label{terrmutau}\\
&&\qquad\qquad L/E\sim0.02.\nonumber
\end{eqnarray}
For the experiments searching for $\nu_{\mu}\to\nu_{e}$ and $\bar{\nu}_{\mu}\to\bar{\nu}_e$ oscillations, we used the experiments E776 \CITE{E776}, KARMEN \cite{EMU} and LSND \cite{LSND} accelerator experiments;
\begin{mathletters}
\label{terrmue}
\begin{eqnarray}
&&P(\nu_{\mu}\to\nu_{e})<3\times10^{-3}\ \ (90\%\ {\rm C.L.}),\ \ {\rm E776}\label{terrmuea}\\
&&\qquad\qquad L=1{\rm km},\ \ \ \ E \sim 1{\rm GeV},\nonumber\\
&&P(\bar{\nu}_{\mu}\to\bar{\nu}_{e})<3.1\times10^{-3}\ \ (90\%\ {\rm C.L.}),\ \ {\rm KARMEN}\label{terrmueb}\\
&&\qquad\qquad L=17.5{\rm m},\ \ \ \ E<50{\rm MeV},\nonumber\\
&&P(\bar{\nu}_{\mu}\to\bar{\nu}_{e})=3.4\mbox{\small${+2.0\atop-1.8}$}\pm0.7\times10^{-3},\ \  {\rm LSND}\label{terrmuec}\\
&&\qquad\qquad  L=30{\rm m},\ \ \ \ E\sim36-60{\rm MeV}.\nonumber
\end{eqnarray}
\end{mathletters}
We analyzed the Bugey experiment using nuclear power reactor \cite{BUGEY} searching for the disappearance of $\bar{\nu}_e$. In this paper, we add the Chooz experiment\cite{CHOOZ}.
\begin{mathletters}
\label{terree}
\begin{eqnarray}
&&1-P(\bar{\nu}_{e}\to\bar{\nu}_{e})<10^{-2}\ \ (90\%\ {\rm C.L.}),\ \ {\rm Bugey}\label{terreea}\\
&&\qquad\qquad L=15,\ 40,\ 95{\rm m},\ \ \ E\sim1-6{\rm MeV}.\nonumber \\
&&1-P(\bar{\nu}_{e}\to\bar{\nu}_{e})<0.9\times10^{-1}\ \ (90\%\ {\rm C.L.}),\ \ {\rm Chooz}\label{terreeb}\\
&&\qquad\qquad L=1{\rm km},\ \ \ E\sim3{\rm MeV}.\nonumber
\end{eqnarray}
\end{mathletters}
\par 
We show the contour plots of the allowed regions on $(\tan^2\theta_{13},\ \tan^2\theta_{23}$) plane determined by the probability $P$ expressed in Eq.~(\ref{trans.prob.}) and above experimental data Eqs.~(\ref{terrmutau}), (\ref{terrmue}) (except LSND data (\ref{terrmuec})) and (\ref{terree}) in Fig.~\ref{fig5}. Allowed regions are left and right hand sides surrounded by curves. Curves represent the boundary of $90$ \% C.L. of $P$. We fixed the parameters $\Delta m_{12}^2$ and $\theta_{12}$ as $\Delta m_{12}^2=10^{-5}{\rm eV}^2$ and $\sin^22\theta_{12}=0.8$, and the parameter $\Delta m_{13}^2$ to be various values from 0.001eV$^2$ to 1eV$^2$. Although we fix the parameters $\Delta m_{12}^2$ and $\theta_{12}$ as $\Delta m_{12}^2=10^{-5}{\rm eV}^2$ and \ $\sin^22\theta_{12}=0.01$, the results are not changed because the probability $P$ of terrestrial neutrino is insensitive to the $e-\mu$ mixing parameters $\Delta m^2_{12}$ and $\sin^22\theta_{12}$. Dotted lines show the allowed regions restricted by the LSND data Eq.\ (\ref{terrmuec}).  From these results, we see that the allowed regions on $(\theta_{13},\ \theta_{23})$ is very restricted by LSND data. Comparing the result obtained previous analysis\cite{TESHIMA} with this, there is observed that on mass parameter $\Delta m^2_{23}\sim 10^{-2}-2\times10^{-3}$, allowed region for $\tan^2\theta_{13}$ in present result is rather restrictive than the one in previous result.  

\subsection{Allowed regions of mixing parameters}
We here discuss the allowed regions for mixing parameters $(\Delta m^2_{12},\ \sin^22\theta_{12},\ \Delta m^2_{23},\ \theta_{13},\\ \theta_{23})$ satisfying atmospheric neutrino and terrestrial neutrino experiments. $\Delta m^2_{12}$ and $\sin^22\theta_{12}$ have been determined in analyses considering MSW effects of the solar neutrino experiments\cite{FOGLIS,TESHIMA}:                                                                   \begin{eqnarray}
(\Delta m^2_{12},&& \sin^22\theta_{12})\nonumber\\
=&&\left\{\begin{array}{l}
(4\times10^{-6}-7\times10^{-5}{\rm eV}^2,\ 0.6-0.9),\ \ \ {\rm large\ angle\ solution}\\
(3\times10^{-6}-1.2\times10^{-5}{\rm eV}^2,\ 0.003-0.01).\ \ \ {\rm small\ angle\ solution}
\end{array}\right.\\
&& \qquad\qquad {\rm for}\ \theta_{13}=0^\circ-20^\circ\nonumber
\end{eqnarray}
\par
Observing the allowed regions obtained in terrestrial neutrino experiments (Fig.  \ref{fig5}) and the allowed regions obtained in atmospheric neutrino experiments including sub-GeV and multi-GeV data but not including zenith angle dependence's (Fig. \ref{fig2}), we obtain the allowed regions satisfying these experiments.
\begin{mathletters}
\begin{eqnarray}
&& \makebox{large angle solution} \qquad\qquad\quad \makebox{small angle solution} \nonumber\\
&{\rm for}\ \Delta m_{23}^2=10{\rm eV}^2\ \ \ \ \ \ \  & \makebox{no allowed region}\qquad\qquad\qquad  \makebox{no allowed region} \\
&{\rm for}\ \Delta m_{23}^2=1{\rm eV}^2 \ \ \ \ \ \ \ \ &  
          (\theta_{13}<4^\circ,\ 24^\circ<\theta_{23}<62^\circ)\ \ \ \ \ 
          (\theta_{13}<4^\circ,\ 24^\circ<\theta_{23}<66^\circ)\\
&{\rm for}\ \Delta m_{23}^2=0.1{\rm eV}^2\ \ \ \ \  &   
          (\theta_{13}<3^\circ,\ 27^\circ<\theta_{23}<63^\circ)\ \ \ \ \   
          (\theta_{13}<3^\circ,\ 27^\circ<\theta_{23}<63^\circ) \\
&{\rm for}\ \Delta m_{23}^2=0.01{\rm eV}^2\ \ \ \ &  
          (\theta_{13}<13^\circ,\ \theta_{23}\sim45^\circ)\qquad\quad \ \ \ 
          (\theta_{13}<13^\circ,\ \theta_{23}\sim45^\circ) \\
&{\rm for}\ \Delta m_{23}^2=0.001{\rm eV}^2\ \ \  & \makebox{no allowed region}\qquad\qquad\qquad  \makebox{no allowed region} 
\end{eqnarray}
\end{mathletters}
\par
We next show the allowed regions satisfying the terrestrial experiment data (Fig. 5) and zenith angle dependence's of atmospheric experiments (Fig. 3). We showed the allowed regions for sub-GeV data and multi-GeV data, separately.\\
\par
For sub-GeV case: allowed region are obtained for large angle solution. 
\begin{mathletters}
\begin{eqnarray}
&{\rm for}\ \Delta m_{23}^2=10{\rm eV}^2\ \ \ & \ \ 
          {\rm no\ allowed\ region}\quad\ \ \  \\
&{\rm for}\ \Delta m_{23}^2=1{\rm eV}^2 \ \ \ \ & \ \    
          (\theta_{13}<4^\circ,\ 18^\circ<\theta_{23}<32^\circ)\quad\ \ \ \\
&{\rm for}\ \Delta m_{23}^2=0.1{\rm eV}^2 \ \ & \ \   
          (\theta_{13}<3^\circ,\ 18^\circ<\theta_{23}<33^\circ)\ \ \ \ \ \ \\
&{\rm for}\  \Delta m_{23}^2=0.01{\rm eV}^2 \ & \ \  
          (\theta_{13}<13^\circ,\ 17^\circ<\theta_{23}<32^\circ)\ \ \ \ \\
&{\rm for}\ \Delta m_{23}^2=0.001{\rm eV}^2 & \ \ 
          (\theta_{13}<23^\circ,\ 21^\circ<\theta_{23}<35^\circ) 
\end{eqnarray}
\end{mathletters}
\par
For multi-GeV case: allowed region are obtained for large and small angle solutions. 
\begin{mathletters}
\begin{eqnarray}
&{\rm for}\ \Delta m_{23}^2=10{\rm eV}^2\ \ \ & \ \ 
          {\rm no\ allowed\ region}\quad\ \ \  \\
&{\rm for}\ \Delta m_{23}^2=1{\rm eV}^2 \ \ \ \ & \ \    
          (\theta_{13}<4^\circ,\ 27^\circ<\theta_{23}<63^\circ)\quad\ \ \ \\
&{\rm for}\ \Delta m_{23}^2=0.1{\rm eV}^2 \ \ \ & \ \   
          (\theta_{13}<3^\circ,\ 28^\circ<\theta_{23}<61^\circ)\ \ \ \ \ \ \\
&{\rm for}\  \Delta m_{23}^2=0.01{\rm eV}^2 \ & \ \  
          (\theta_{13}<13^\circ,\ 35^\circ<\theta_{23}<55^\circ)\ \ \ \ \\
&{\rm for}\ \Delta m_{23}^2=0.001{\rm eV}^2 & \ \ 
          (11^\circ<\theta_{13}<23^\circ,\ 45^\circ<\theta_{23}<61^\circ) 
\end{eqnarray}
\end{mathletters}
\par 
We finally show the allowed regions satisfying the terrestrial experimental data (Fig. 5) and atmospheric experiments including the zenith angle dependence's for both sub-GeV and multi-GeV data (Fig. 4).\\
\par
For sub-GeV plus multi-GeV case: allowed region are obtained for large angle solution and $\Delta m^2_{13}=0.08-2{\rm eV}^2$. 
\begin{mathletters}
\begin{eqnarray}
&{\rm for}\ \ \Delta m_{23}^2=2{\rm eV}^2\ \ \ & \ \ 
          (\theta_{13}<4^\circ,\ \theta_{23}\sim30^\circ)\quad\ \ \ \\
&{\rm for}\ \ \Delta m_{23}^2=1{\rm eV}^2 \ \ \ \ & \ \    
          (\theta_{13}<4^\circ,\ 27^\circ<\theta_{23}<32^\circ)\quad\ \ \ \\
&{\rm for}\ \ \Delta m_{23}^2=0.1{\rm eV}^2 \ \ & \ \   
          (\theta_{13}<3^\circ,\ 28^\circ<\theta_{23}<33^\circ)\ \ \ \ \ \ \\
&{\rm for}\ \ \Delta m_{23}^2=0.08{\rm eV}^2 \ & \ \  
          (\theta_{13}<3^\circ,\ \theta_{23}\sim30^\circ)\quad\ \ \   
\end{eqnarray}
\end{mathletters}
These regions are very restricted than atmospheric experiments regions Eqs. (15) obtained by not considering the zenith angle dependence's. Although the LSND data is very restrictive experiment, there exists allowed region satisfying this experiment at the values of parameters $(\Delta m^2_{23}\sim1{\rm eV}^2,\ \theta_{13}=3-4^\circ,\ \theta_{23}=27-32^\circ)$ with large angle solution. 

\section{Discussions}
We analyzed atmospheric neutrino experiments of SuperKamiokande\cite{SUPERKAMIOKANDE,KANEYUKI} using the three-flavor neutrino framework with the mass hierarchy $m_1\approx m_2\ll m_3$ and obtained the allowed regions of the parameters $(\Delta m_{12}^2,\ \sin^22\theta_{12},\ \Delta m_{23}^2,\ \theta_{13},\ \theta_{23})$. We also analyzed the terrestrial neutrino experiments including Chooz data\cite{CHOOZ}. From these neutrino experiments (not considering the zenith angle dependence) and the results obtained in solar neutrino experiments\cite{TESHIMA}, we obtained the rather wide allowed regions denoted in Eqs. (15). When zenith angle dependence's are considered, the allowed regions are restricted as follows: $(\Delta m_{12}^2=4\times10^{-6}-7\times10^{-5}{\rm eV}^2,\ \sin^22\theta_{12}=0.6-0.9,\ \Delta m^2_{23}=0.08-2{\rm eV}^2)$, and $(\theta_{13}<4^\circ,\ 27^\circ<\theta_{23}<32^\circ)$ for $\Delta m_{23}^2=1{\rm eV}^2$, $(\theta_{13}<3^\circ,\ 28^\circ<\theta_{23}<33^\circ)$ for $\Delta m_{23}^2=0.1{\rm eV}^2$. If we include LSND experiment in terrestrial experiments, allowed regions are restricted to $(\Delta m_{23}^2\sim1{\rm eV}^2,\ \theta_{13}=3-4^\circ,\ \theta_{23}=27-32^\circ)$.
\par
Finally, we present the neutrino mixing matrix Eq. (2) numerically for the allowed solutions Eqs. (18);
\begin{equation}
U=\left(\begin{array}{ccc}
       0.81 \leftrightarrow 0.90 & 0.43 \leftrightarrow 0.58 & 0.0 \leftrightarrow 0.07\\
       -0.36 \leftrightarrow -0.55 & 0.66 \leftrightarrow 0.80 & 0.45 \leftrightarrow 0.54\\
       0.14 \leftrightarrow 0.32 & -0.37 \leftrightarrow -0.51 & 0.84 \leftrightarrow 0.89
       \end{array}\right).
\end{equation}
         

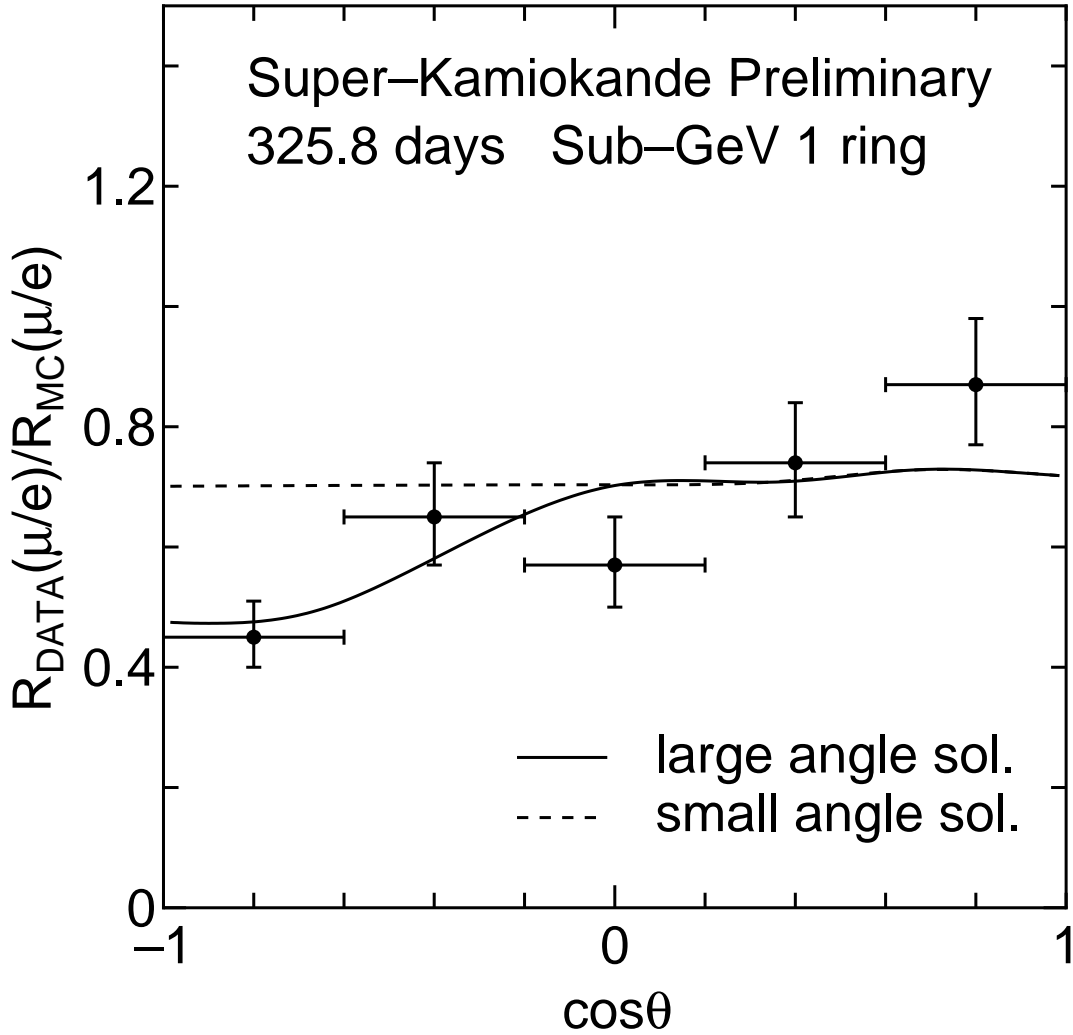
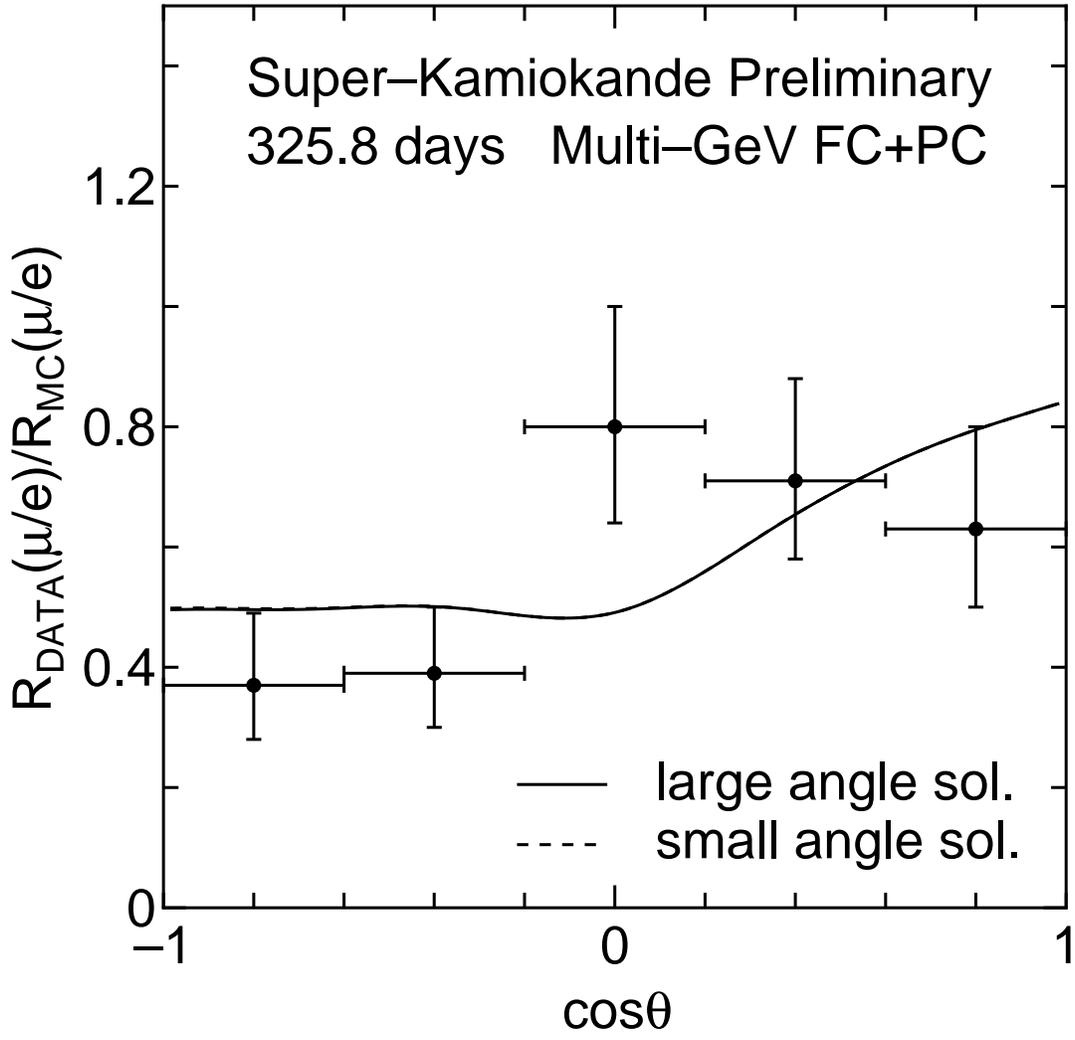
\begin{figure}
\caption{The zenith angle dependence of $R(\mu/e)$ in atmospheric neutrino experiment. Experimental data is referred to SuperKamiokande (Ref. [2]). Fig. (a) represents the sub-GeV experimental case: solid curve is calculated on typical parameters for large angle solution $(\Delta m_{12}^2=$ $3\times10^{-5}{\rm eV}^2,$ $\ \sin^22\theta_{12}=0.7,\ \Delta m_{23}=0.2{\rm eV}^2,\ \theta_{13}=4^\circ,\ \theta_{23}=25^\circ)$ and dotted curve on small angle solution $(\Delta m_{12}^2=10^{-5}{\rm eV}^2,\ \sin^22\theta_{12}=0.007,\ \Delta m_{23}=0.2{\rm eV}^2,$ $\ \theta_{13}=4^\circ,$ $\ \theta_{23}=25^\circ)$. Fig. (b) represents the multi-GeV experimental case: solid curve is calculated on parameters $(\theta_{13}=4^\circ,\ \theta_{23}=30^\circ)$ for large angle solution  and dotted curve on parameters $(\theta_{13}=4^\circ,$ $\ \theta_{23}=30^\circ)$ for small angle solution.}
\label{fig1}
\end{figure}

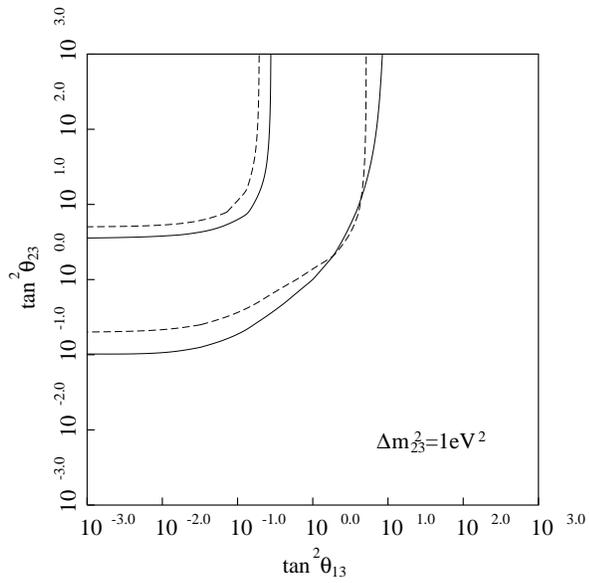
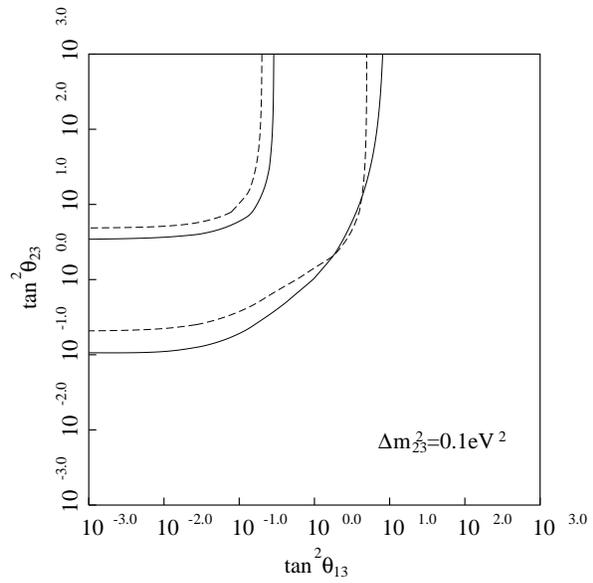
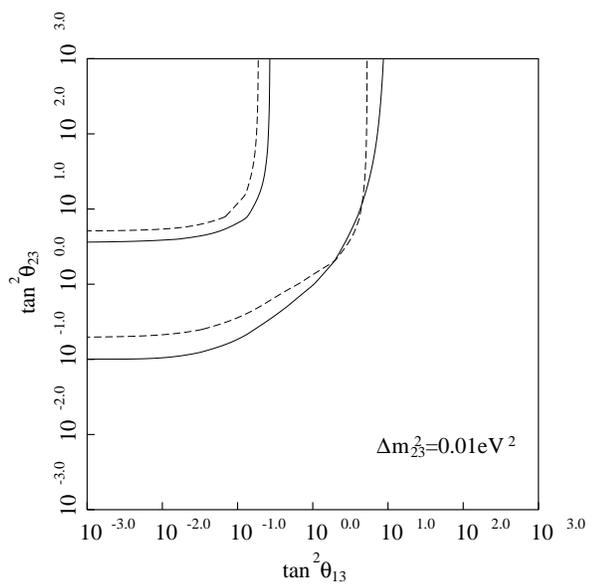
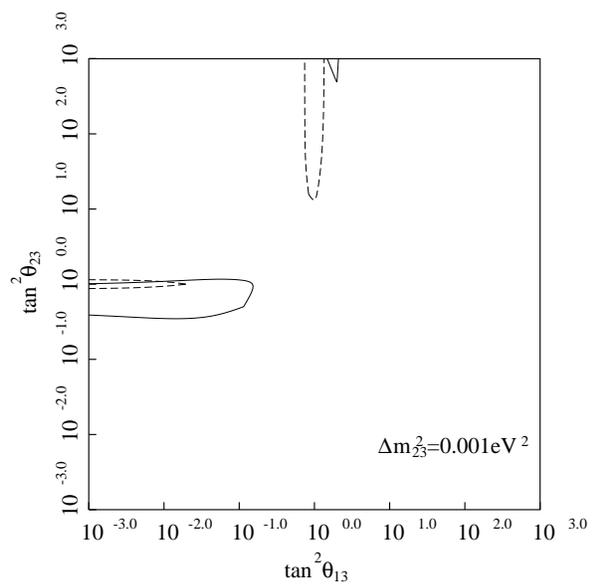
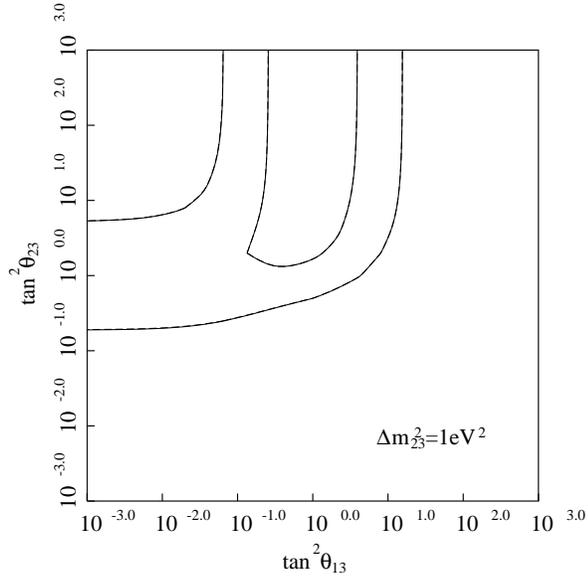
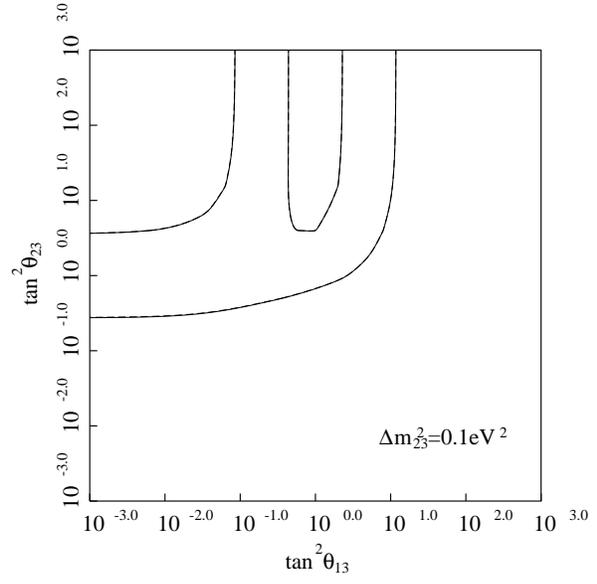
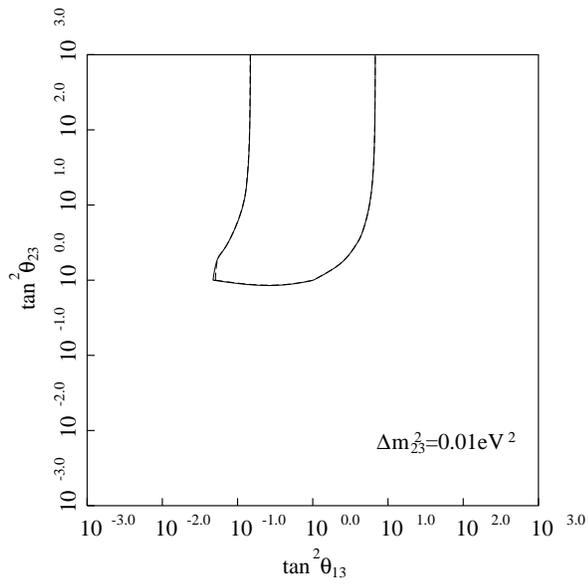
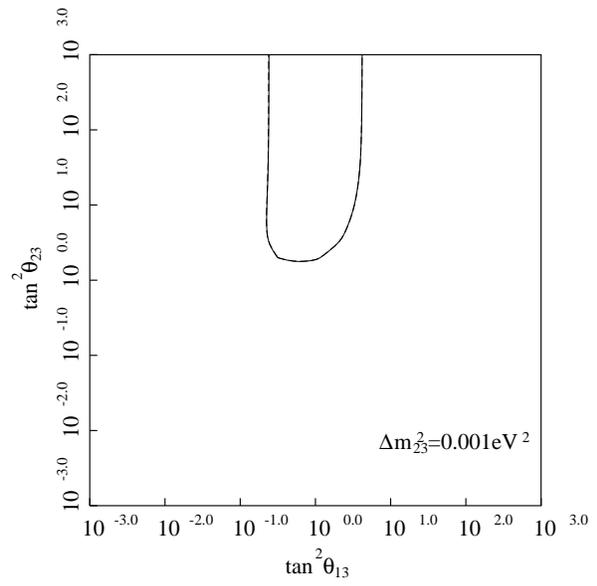
\begin{figure}
\caption{The plots of allowed regions determined by atmospheric neutrino data Eq. (9c) of $R(\mu/e)$ for various values of $\Delta m_{23}^2 $ on $\tan^2\theta_{13}-\tan^2\theta_{23}$ .Figs. (a)-(d) show the plots of sub-GeV experiments and Figs. (e)-(h) the ones of multi-GeV experiments. We fix the parameters $(\Delta m_{12}^2$, $\sin^22\theta)$ to be $(3\times10^{-5}{\rm eV}^2,\ 0.7)$ corresponding to the large angle solution (solid lines) and  $(10^{-5}{\rm eV}^2,\ 0.005)$ corresponding to the small angle solution (dotted lines).} 
\label{fig2}
\end{figure}

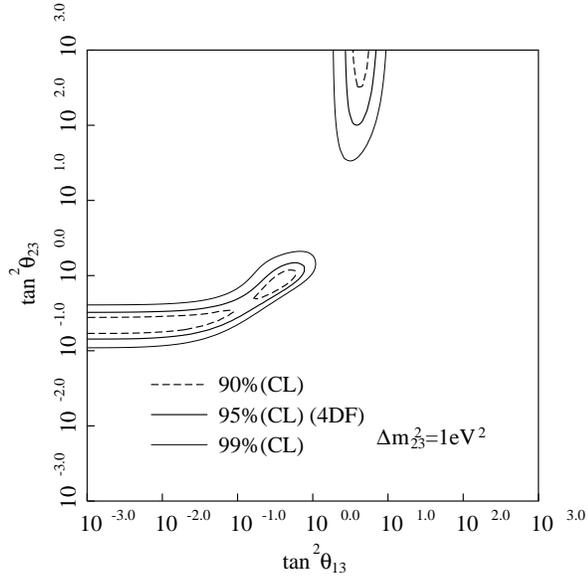
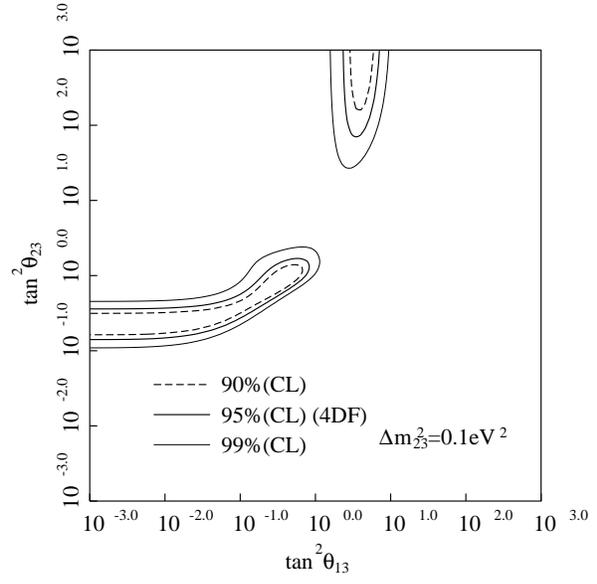
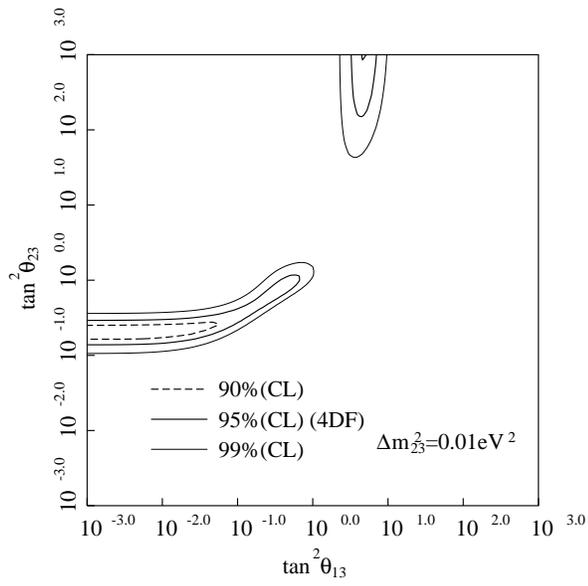
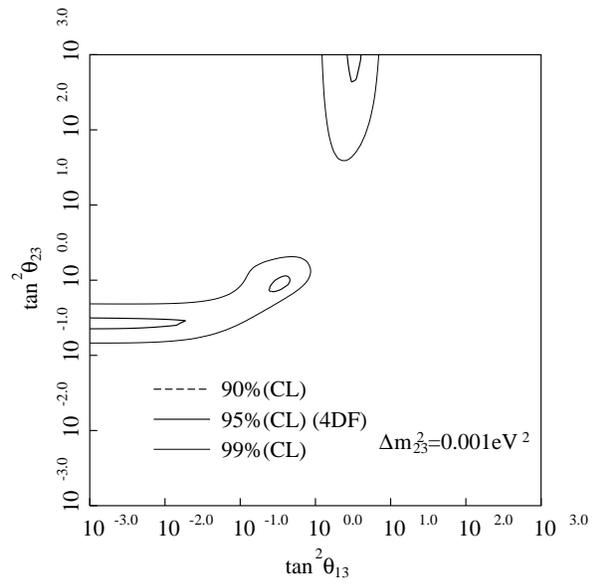
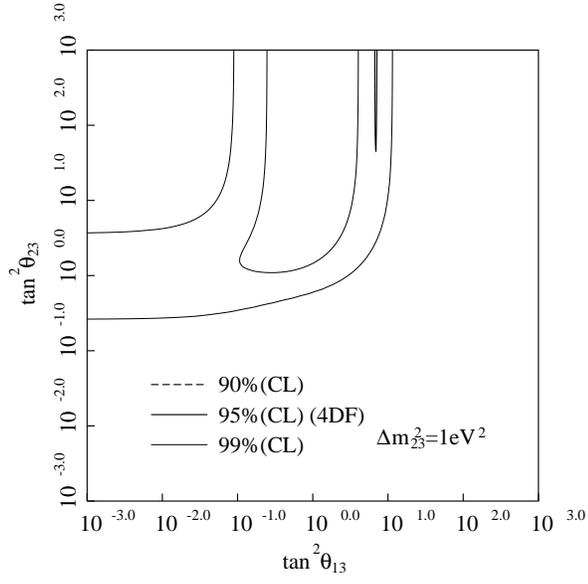
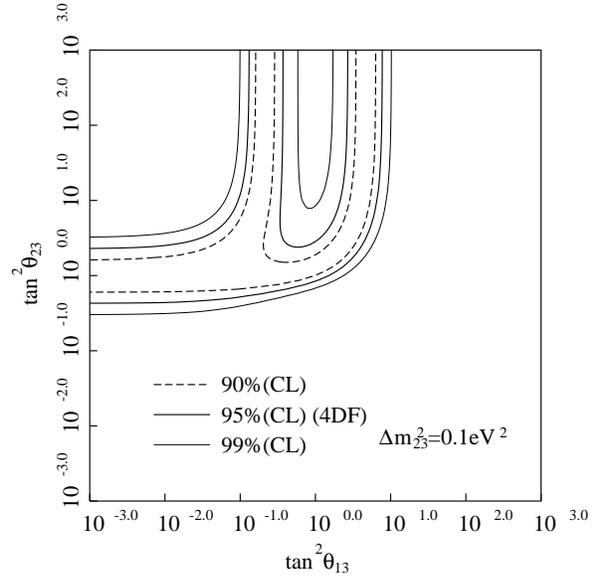
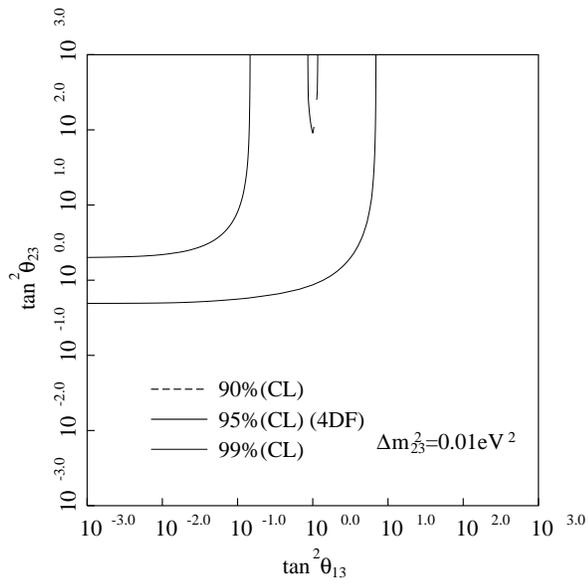
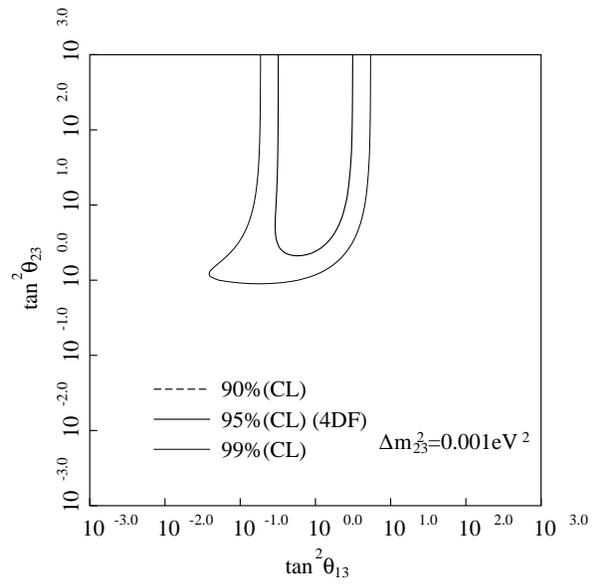
\begin{figure}
\caption{ The plots of allowed regions on $\tan^2\theta_{13}-\tan^2\theta_{23}$ plane determined by the zenith angle dependence's of SuperKamiokande data using the $\chi^2$ analysis.  Figs. (a)-(d) shows the plots of sub-GeV experiments for large angle solution of $\sin^22\theta_{12}$ and Figs. (e)-(h) the ones of multi-GeV experiments for large and small angle solutions. Solid thin, solid thick and dotted curves denote the regions allowed in $99\%$, $95\%$ and $90\%$ C.L., respectively.}
\label{fig3}
\end{figure}

\begin{figure}
\caption{ The plots of allowed regions on $\tan^2\theta_{13}-\tan^2\theta_{23}$ plane determined by the sub-GeV and multi-GeV zenith angle dependence's of SuperKamiokande data using the $\chi^2$ analysis. These are taken for large angle solution. Solid thin, solid thick and dotted curves denote the regions allowed in $99\%$, $95\%$ and $90\%$ C.L., respectively. }
\label{fig4}
\end{figure}

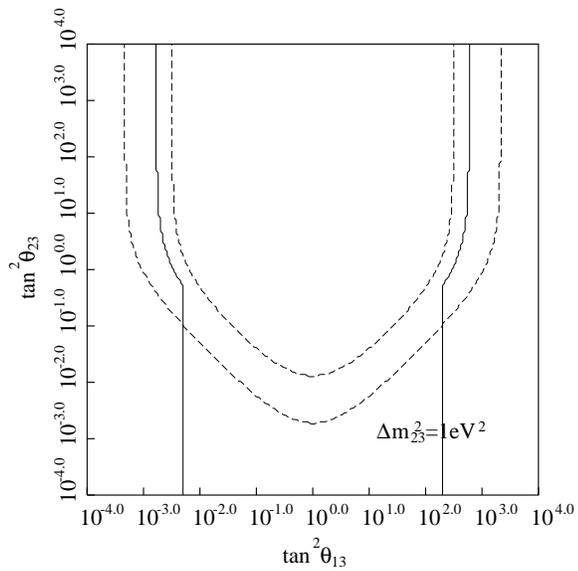
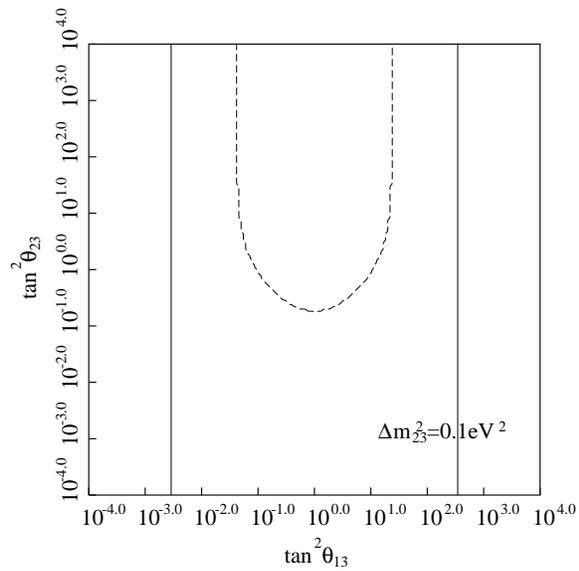
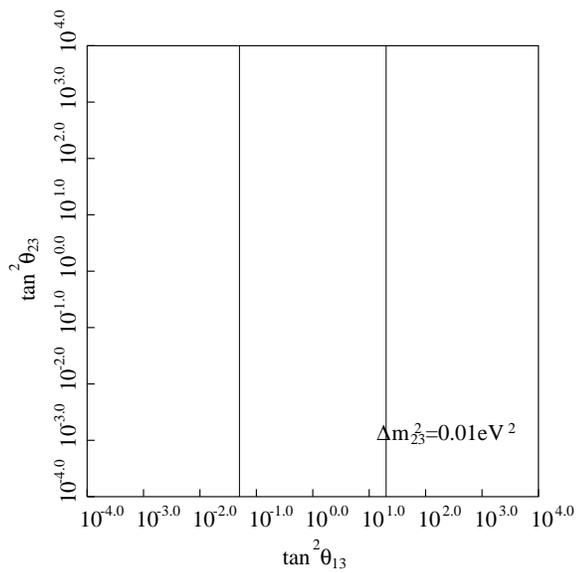
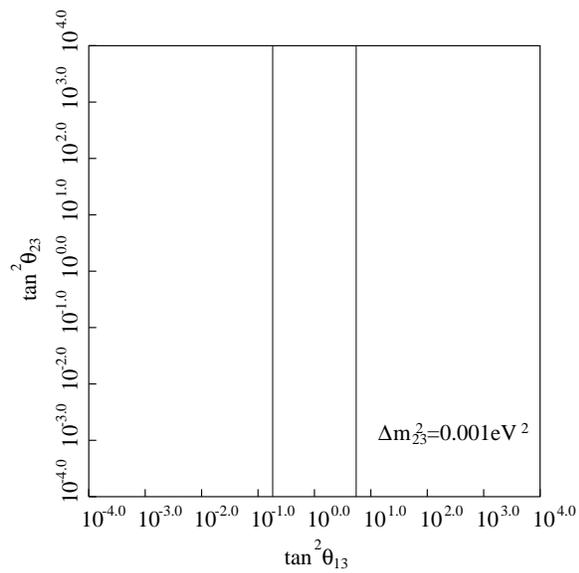
\begin{figure}
\caption{The plots of allowed regions on $\tan^2\theta_{13}-\tan^2\theta_{23}$ plane determined by $P$ of terrestrial $\nu_{\mu}\to\nu_{\tau}$, ${\nu}_{\mu}\to\nu_{e}$, $\bar{\nu}_{\mu}\to\bar{\nu}_{e}$ and $\bar{\nu}_e\to\bar{\nu}_e$ experiments. Allowed regions are left and right hand sides surround by curves and/or  lines. Curves and lines represent the boundary of $90$ \% C.L. of $P$. $\Delta m_{12}^2$ and $\sin^22\theta_{12}$ are fixed as $10^{-5}{\rm eV}^2$ and 0.8, respectively. $\Delta m_{23}^2$ is fixed to 1eV$^2$(Fig.~5(a)), 0.1eV$^2$(Fig.~5(b)), 0.01eV$^2$(Fig.~5(c)) and 0.001eV$^2$(Fig.~5(d)). Dotted lines show the allowed regions determined by LSND data.} 
\label{fig5}
\end{figure}


\begin{thebibliography}{99}
\bibitem{SUPERKAMIOKANDE} Y.~Totsuka(SuperKamiokande Collaboration), in {\it LP'97}, 28th International Symposium on Lepton Photon Interactions, Hamburg, Germany, 1997, to appear in the Proceedings.
\bibitem{KANEYUKI} K. Kaneyuki(SuperKamiokande Collaboration), in {\it 1997 Sectional Meetings of the Physical Society of Japan}, Tokyo Metropolitan Univ., Japan, 1997.

\bibitem{HIRATA} Kamiokande Collaboration, K.~S.~Hirata {\it et al.}, Phys. Lett. {\bf B280}, 146(1992). 
\bibitem{FUKUDA} Kamiokande Collaboration, Y.~Fukuda {\it et al.}, Phys. Lett. {\bf B335}, 237(1994).
\bibitem{E531} E531 Collaboration, N.~Ushida {\it et al.}, Phys. Rev. Lett. {\bf57}, 2897(1986). 
\bibitem{CHORUS} K.~Winter, Nucl. Phys. {\bf B}(Proc. Suppl.){\bf38}, 211(1995).
\bibitem{E776}  E776 Collaboration, L.~Borodovsky {\it et al.}, Phys. Rev. Lett. {\bf 68}, 274(1992).
\bibitem{EMU} KARMEN Collaboration, B.~Armbruster {\it et al.}, Nucl. Phys. {\bf B}(Proc. Suppl.){\bf38}, 235(1995).
\bibitem{LSND} LSND Collaboration, C. Athanassopoulos {\it et al.}, Phys. Rev. Lett. {\bf 75}, 2650 (1995). J. E. Hill, Phys. Rev. Lett. {\bf 75}, 2654(1995).  W.~C.~Louis, Nucl. Phys. {\bf B}(Proc. Suppl.){\bf38}, 229(1995).
\bibitem{BUGEY} B.~Achkar {\it et al.}, Nucl. Phys. {\bf B434}, 503(1995).
\bibitem{CHOOZ} M. Apollonio {\it et al}, hep-ex/9711002.
\bibitem{TESHIMA} T. Teshima, T. Sakai and O. Inagaki, hep-ph/9801276.
\bibitem{KUO} T.~Kuo and J.~Pantaleone, Rev.~Mod.~Phys. {\bf 61}, 937(1989).
\bibitem{FOGLIS} G.~L.~Fogli, E.~Lisi and D.~Montanino, Phys. Rev. {\bf D54}, 2048(1996).
\bibitem{VACUUME} P. I. Krastev and S. T. Petcov, Phys. Rev. Lett. {\bf 72}, 1960(1996). 
\bibitem{FOGLIT}  G. L. Fogli, E. Lisi and G. Sciscia, Phys.~Rev.~{\bf D52}, 5334(1995). G. L. Fogli, E. Lisi and G. Sciscia, Phys.~Rev.~{\bf D 56}, 3081(1997).
\bibitem{FOGLIA}   G.~L.~Fogli, E.~Lisi and D.~Montanino, Phys. Rev. {\bf D52}, 2775(1995). G.L. Fogli, E. Lisi, D. Montanio and G. Sciscia, Phys.~Rev.~{\bf D55}, 4385(1997).  C. Giunti, C. W. Kim and M. Monteno, hep-ph/9709439.
\bibitem{THREE}  C. Y. Cardall and G. M. Fuller, Phys.~Rev.~{\bf D 53}, 4421(1996). G. L. Fogli, E. Lisi, D. Montanio and G. Sciscia, Phys.~Rev.~{\bf D 56}, 4365(1997). 
\bibitem{KAMIOKANDE2} Y. Suzuki, in {\it Proceedings of 28th International Conference on High Energy Physics}, Warsaw, Poland, 1996, edited by Z. Ajduk and A. K. Wroblewski (Warsaw University), p.273.
\bibitem{IMB} IMB Collaboration, R. Becker-Szendy {\it et al.}, Phys. Rev. {\bf D46}, 3720(1992).
\bibitem{SAGE} SAGE Collaboration, J.~N.~Abdurashitov {\it et al.}, Phys.~Lett.~{\bf B328}, 234(1994).
\bibitem{GALLEX} GALLEX Collaboration, P.~Anselmann {\it et al.}, Phys.~Lett.~{\bf B357}, 237(1995).
\bibitem{HOMESTAKE} B.~T.~Cleveland {\it et al.}, Nucl.~Phys.~{\bf B}(Proc.~Suppl.){\bf 38}, 47(1995).
\bibitem{KAMIOKANDES} Kamiokande Collaboration, K.~S.~Hirata {\it et al.}, Phys.~Rev.~{\bf D44}, 2241(1991); Phys.~Rev.~Lett.~{\bf 66}, 9(1991). 
\bibitem{GAISSER} T. K. Gaisser, T. Stanev and G. Barr, Phys. Rev. {\bf D38}, 85(1988).
\bibitem{HONDA} M. Honda {\it et al.}, Phys. Rev. {\bf D52}, 4985(1995).
\bibitem{AGRAWAL} Vivek Agrawal {\it et al.}, Phys. Rev. {\bf D53}, 1314(1996).
\end{thebibliography}
\end{document}